\documentclass[pre,twocolumn,longbibliography]{revtex4}
\usepackage[cp850]{inputenc}

\usepackage{amssymb,amsmath}

\usepackage{verbatim}

\newcommand\beq{\begin{equation}}
\newcommand\eeq{\end{equation}}
\newcommand\beqa{\begin{eqnarray}}
\newcommand\eeqa{\end{eqnarray}}
\newcommand{\al}{\alpha}

\newcommand*\Bell{\ensuremath{\boldsymbol\ell}}

\usepackage{color}

\newcommand{\vicente}[1]{{ #1}}

\usepackage{graphicx}
\usepackage{epstopdf}

\begin{document}

\title{On the mean square displacement of intruders in freely cooling granular gases}
\author{Enrique Abad}
\affiliation{Departamento de F\'{\i}sica Aplicada and Instituto de Computaci\'on Cient\'{\i}fica Avanzada (ICCAEx), Universidad de Extremadura, E-06800 M\'erida, Spain}
\author{Santos Bravo Yuste and Vicente Garz\'o\email{vicenteg@unex.es}}
\affiliation{Departamento de F\'{\i}sica and Instituto de Computaci\'on Cient\'{\i}fica Avanzada (ICCAEx), Universidad de Extremadura, E-06071 Badajoz, Spain}

\begin{abstract}

We compute the mean square displacement (MSD) of intruders immersed in a freely cooling granular gas made up of smooth inelastic hard spheres. In general, intruders and particles of the granular gas are assumed to have different mechanical properties, implying that non-equipartition of energy must be accounted for in the computation of the diffusion coefficient $D$. In the hydrodynamic regime, the time decay of the granular temperature $T$ of the cooling granular gas is known to be dictated by Haff's law; the corresponding decay of the intruder's collision frequency entails a time decrease of the diffusion coefficient $D$. Explicit knowledge of this time dependence allows us to determine the MSD by integrating the corresponding diffusion equation. As in previous studies of self-diffusion (intruders mechanically equivalent to gas particles) and the Brownian limit (intruder's mass much larger than the grain's mass), we find a logarithmic time dependence of the MSD as a consequence of Haff's law. This dependence extends well beyond the two aforementioned cases, as it holds in all spatial dimensions for arbitrary values of the mechanical parameters of the system (masses and diameters of intruders and grains, as well as their coefficients of normal restitution). Our result for self-diffusion in a three-dimensional granular gas agrees qualitatively, but not quantitatively, with that recently obtained by Blumenfeld [arXiv: 2111.06260] in the framework of a random walk model. Beyond the logarithmic time growth, we find that the MSD depends on the mechanical system parameters in a highly complex way. We carry out a comprehensive analysis from which interesting features emerge, such a non-monotonic dependence of the MSD on the coefficients of normal restitution and on the intruder-grain mass ratio. To explain the observed behaviour, we analyze in detail the intruder's random walk, consisting of ballistic displacements interrupted by anisotropic deflections caused by the collisions with the hard spheres. We also show that the MSD can be thought of as arising from an equivalent random walk with isotropic, uncorrelated steps.
Finally, we derive some results for the MSD of an intruder
immersed in a driven granular gas and compare them with those obtained
for the freely cooling case. In general, we find significant quantitative differences in the dependence of the scaled diffusion coefficient on the coefficient of normal restitution for the grain-grain collisions.

\end{abstract}

\draft
\date{\today}
\maketitle

\section{Introduction}
\label{sec0}

Everyday life examples of granular media include both monodisperse and polydisperse systems composed of macroscopic particles (e.g. sand, rice, coffee, etc.). However, the physics of granular materials still holds on much larger
spatial scales, e.g., those typical of astrophysical systems such as interstellar clouds \cite{WP93}, populations of asteroids \cite{HSS19}, or planetary rings \cite{GB84}. This diversity of sizes makes granular matter of great interest in many different contexts. In particular, granular media provide the opportunity of studying diffusive transport at macroscopic scale \cite{DLDC12,LBDG18,LGRAYV21}.  At such a scale, thermal fluctuations -the intrinsic source of stochasticity in mesoscopic diffusive systems-, do not play any significant role. This irrelevance of thermal noise facilitates the tuning of the different system parameters and the ultimate control of typical transport characteristics such as diffusivities and diffusion exponents \cite{LGRAYV21}.

The dissipative character of the collisions suffered by the elementary units of granular matter imply that their diffusive motion will eventually stop in the absence of an external energy input. Therefore, studies of diffusion in such inherently non-equilibrium systems are often carried out in driven steady states exhibiting sustained particle motion \cite{IALOB52,OU99,WHP01,WP02,MVPRKEU05}. However, diffusion can also be studied in freely (undriven) cooling systems, well before major disruptions like clustering instabilities start to occur \cite{GZ93,M93}. Relaxation phenomena in this so-called homogeneous cooling state (HCS) continue to pose great theoretical and practical challenges for researchers. In this context, the detailed characterization of diffusive transport both for monocomponent and multicomponent systems is an important aspect with some as yet unsolved questions.

A deceivingly simple object of study is the HCS of a monocomponent granular gas of inelastic smooth hard spheres. Here, the inelasticity in collisions is only accounted for by the (positive) coefficient of normal restitution $\alpha$. This quantity measures the ratio between the magnitude of the normal component of the relative velocity (oriented along the line separating the centers of the two spheres at contact)  before and after a collision.  In the simplest case, the coefficient $\al$ is taken to be constant. Despite the apparent simplicity of the HCS, the exact analytic form of its time-dependent velocity distribution function is not known to date. An intruder moving inside one such cooling granular gas collides inelastically with the surrounding particles, and it consequently loses kinetic energy in the course of time. This energy loss entails a drastic slowing-down of the intruder's motion, to the extent that its diffusion becomes \emph{ultraslow} \cite{MJChB14,BChM15a,LWCZM19}; in other words, the mean square displacement (MSD) is of the form $\langle |\Delta \mathbf{r}|^2(t)\rangle \propto [\ln(c_1+c_2 t)]^\eta$, thus exhibiting a time decay slower than any power law. The quantities $c_1$ and $c_2$ depend on the mechanical properties of the system. For the above HCS system, we will see that one has $\eta=1$. The mechanism leading to ultraslow diffusion here is fundamentally different from those present in other systems with diverse values of $\eta$; in those systems, the slowing-down of the motion arises notably from crowding \cite{BO02}, energy trapping \cite{CKS03}, spatial disorder \cite{S82}, a combination of the latter with many-body interactions \cite{SLLFMA14}, or the interplay between a harmonic force and a medium that expands at a suitably chosen rate \cite{LVYA19}.

To the best of our knowledge, the study of the intruder's MSD has focused on two limit cases, i.e., self-diffusion (intruder mechanically equivalent to gas particles) and Brownian limit (intruder's mass much larger than the mass of the gas particles). For self-diffusion in gases of smooth inelastic hard spheres with constant $\alpha$, previous studies \cite{BRCG00,BP00,BChM15b} have shown that Haff's law leads to a diffusion coefficient whose long-time decay is inversely proportional to time, resulting in the aforementioned logarithmic time growth of the MSD. For such systems, Brey \emph{et al.} \cite{BRCG00} performed a comparison between theoretical results for the self-diffusion coefficient in the first Sonine approximation and DSMC simulations; they found excellent agreement even for moderately strong inelasticity.  At about the same time, Brilliantov and P\"oschel \cite{BP00} also mentioned that a logarithmic time dependence of the MSD arises in the self-diffusive case without writing this quantity in full detail.  More recently, Bodrova \emph{et al.} \cite{BChM15b} gave explicitly the expression for the MSD of a given particle in a three-dimensional gas of hard spheres.

In the Brownian limit, a purely logarithmic time growth of the intruder's MSD is also observed after a short transient regime \cite{BRGD99}. The ultimate reason for this not entirely intuitive finding is once again Haff's law, which only refers to the mechanical properties of the granular gas.

The above findings strongly suggest that the logarithmic behaviour of MSD found in the HCS system considered above holds beyond the specific cases of self-diffusion and the Brownian limit, and that it is therefore insensitive to the details of the intruder's mechanical properties as long as $\alpha$ remains constant (the logarithmic time dependence breaks down for viscoelastic gases with velocity-dependent restitution coefficients, see e.g. Refs.\ \cite{BChM15b} and \cite{BDPB02}). The theoretical verification of this conjecture and the derivation of an explicit expression for the MSD via a Boltzmann-Enskog approach is one of the goals of the present work. However, we will go well beyond this first objective in that (i) we will compare our results to those recently obtained by Blumenfeld \cite{B21} in the framework of a random walk model for the self-diffusive case in three dimensions, (ii) we will perform a comprehensive study of the dependence of the MSD on the different system parameters in the general case, and (iii) we will provide physical explanations for the observed behaviour. \vicente{A pivotal element in such explanations is the idea that the main properties of the intruder's motion (viewed as an isotropic random walk) are intimately connected with the way in which the granular gas cools down. In fact, Blumenfeld showed in a straightforward fashion that a minimal random walk model leading to a logarithmically increasing MSD naturally yields the qualitative time decrease prescribed by Haff's law for the mean kinetic energy of the granular gas \cite{B21}. Our goal here is to enrich this view with some missing quantitative details and to extend it to more general situations. Such an endeavour calls for a rigorous, overarching framework relating kinetic theory to the random walk approach.  We therefore hope that the present paper (along with previous works such as Ref.\ \cite{B21}) will provide additional motivation for further exploring the connections between these two approaches.}

The remainder of this paper is organized as follows. In Sec. \ref{sec1}, we perform kinetic theory calculations to find a formula for the MSD in terms of the mechanical properties of the intruder and of the gas particles. In Sec. \ref{sec2}, we obtain explicit expressions for the MSD in the first-Sonine approximation and show that they are consistent with previous results for the elastic case, the self-diffusive case, and the Brownian limit. Section \ref{Sec-RW} provides a random walk description of the intruder's motion. Section \ref{sec3} is devoted to the self-diffusive case; a significant part of this section is devoted to a detailed comparison between our results and those obtained in the framework of Blumenfeld's approach. In particular, for a fixed, sufficiently large time, we find a  non-monotonic  $\alpha$-dependence of the MSD, which we explain with physical arguments based on our random walk approach. Section \ref{sec4} addresses the general case, in which one no longer has energy equipartition between the intruder and the granular gas particles (in what follows we will also refer to the latter as ``hard spheres'' or ``grains''). In this case, we will show that the MSD displays a non-monotonic dependence on the intruder-grain mass ratio. \vicente{In Sec. \ref{driven}, we derive some results for the MSD of an intruder inmersed in a \emph{driven} granular gas and discuss similarities and differences with respect to the HCS. } Finally, in Sec. \ref{Conc} we summarize our main conclusions and discuss some possible implications for experiments.

\section{Diffusion equation. Mean square displacement}
\label{sec1}

We consider a freely cooling granular gas of mass $m$, diameter $\sigma$, and (constant) coefficient of normal restitution $\alpha$. As said in the Introduction, the coefficient $\al$ measures the ratio between the magnitude of the \emph{normal} component (along the line of separation between the centers of the two spheres at contact) of the pre- and post-collisional relative velocities of the colliding spheres \cite{BP04}. We assume that the granular gas (modeled as a gas of smooth \emph{inelastic} hard spheres) is in the HCS. In this state, the number density $n$ and the granular temperature $T$ are homogeneous; the mean flow velocity $\mathbf{U}$ vanishes, but the granular temperature $T(t)$ decreases in time due to the fact that the binary collisions between grains become inelastic as soon as $\al<1$.

According to the kinetic theory of gases, all the relevant information on the state of the system is provided by the one-particle velocity distribution function $f(\mathbf{v};t)$. For moderately dense gases, the  distribution function $f$ obeys the (inelastic) Enskog--Boltzmann kinetic equation (for homogeneous states, except for the presence of the pair correlation function, the Enskog equation is identical to the Boltzmann equation). From this kinetic equation, the time evolution of the granular temperature
\beq
\label{0}
T(t)=\frac{1}{d n}\int d\mathbf{v}\; m \mbox{v}^2\; f(\mathbf{v};t)
\eeq
can be derived. It is given by \cite{BP04}
\beq
\label{1}
\frac{\partial T}{\partial t}=-\zeta(t) T(t).
\eeq
In Eqs.\ \eqref{0} and \eqref{1}, $d$ is the dimensionality of the system ($d=2$ for hard disks and $d=3$ for hard spheres),
\beq
\label{1.1}
n=\int d\mathbf{v} f(\mathbf{v};t)
\eeq
is the number density of granular gas particles, and
\beq
\label{1.2}
\zeta(t)=-\frac{1}{d n T(t)}\int d\mathbf{v}\; m \mbox{v}^2\, J[f,f]
\eeq
is the cooling rate. Here, $J[f,f]$ is the (inelastic) Enskog--Boltzmann collision operator, whose explicit form is well-known (see, e.g. Ref.\ \cite{BP04}). The cooling rate $\zeta$ provides the rate of energy dissipation due to inelastic collisions, and turns out to be proportional to $1-\alpha^2$ \cite{BP04}. Although the explicit computation of $\zeta$ requires knowledge of the one-particle velocity distribution function $f(\mathbf{v};t)$, dimensional analysis shows that $\zeta(t)\propto \sqrt{T(t)}$, and so the integration of Eq.\ \eqref{1} can be easily performed. The result is
\beq
\label{2}
T(t)=\frac{T(0)}{\left(1+\frac{1}{2}\zeta(0)t\right)^2},
\eeq
where $T(0)$ is the initial temperature and $\zeta(0)$ denotes the cooling rate at $t=0$.


\vicente{Equation \eqref{2} is known as Haff's cooling law for the HCS gas \cite{H83}. Of course, Haff's law holds only as long as the system remains homogeneous; note, however, that a well known characteristic of freely evolving granular gases is the spontaneous formation of velocity vortices and density clusters \cite{GZ93,M93}.  Such instabilities arise from the inelastic character of the collisions, and their main features are well captured by a linear stability analysis of the corresponding hydrodynamic equations. As it turns out,  a critical length $L_c$ (usually associated with the vortex instability) emerges from the analysis, implying that the system becomes unstable as soon as its linear size exceeds $L_c$ \cite{BDKS98,G05,MDCPH11,MGH14,FH17}. Conversely, in small enough systems, the aforementioned instability is suppressed. More precisely, for the particular case of dilute granular gases, the values of $L_c$ [in units of the mean free path $\ell$, see Eq.\ \eqref{9.1} below] for $\alpha=0.9$, 0.8, and 0.7 are $25.3$, $18.8$, and $16.18$, respectively \cite{G05}. Thus,  in order to ensure that the system's homogeneity is preserved, one must limit oneself to time scales on which the MSD remains smaller than $L_c^2$.}

Let us now assume that an impurity or intruder of mass $m_0$ and diameter $\sigma_0$ is added to the granular gas. The presence of the intruder does not have any effect on the state of the granular gas, and thus the HCS is still preserved. Let us denote by $\al_0$ the coefficient of restitution for intruder-gas collisions. Since the intruder and the gas particles are in general mechanically different, one has $\al\neq \al_0$.
The intruder may freely lose or gain momentum and energy in its interactions with the gas particles; therefore, these quantities are not invariants of the (inelastic) Boltzmann--Lorentz collision operator \cite{BP04}. Only the number density of intruders $n_0$ is conserved. This hydrodynamic field is defined as
\beq
\label{2.1}
n_0(\mathbf{r};t)=\int d\mathbf{v} f_0(\mathbf{r}, \mathbf{v};t),
\eeq
where $f_0(\mathbf{r}, \mathbf{v};t)$ is the one-particle velocity distribution function of the intruder. The continuity equation for $n_0$ can be easily obtained from the Boltzmann--Lorentz kinetic equation. It is given by\footnote{Note that, for notational convenience, we define here the flux and diffusion coefficient of the intruder particles in terms of their concentration and not of their mass, as opposed to the definitions used in Ref.\ \cite{G19}.} \cite{G19}
\beq
\label{3}
\frac{\partial n_0}{\partial t}=-\nabla \cdot \mathbf{j}_0,
\eeq
where
\beq
\label{4}
\mathbf{j}_0(\mathbf{r}; t)= \int d\mathbf{v}\; \mathbf{v}\; f_0(\mathbf{r}, \mathbf{v};t)
\eeq
is the intruder particle flux.

The expression for $\mathbf{j}_0$ can be obtained by solving the Boltzmann--Lorentz kinetic equation with the Chapman--Enskog method \cite{CC70}. To first order in $\nabla n_0$, the constitutive equation for $\mathbf{j}_0$ is
\beq
\label{5}
\mathbf{j}_0=-D \nabla n_0,
\eeq
where $D(t)$ is the diffusion coefficient. Substitution of Eq.\ \eqref{5} into Eq.\ \eqref{3} yields
\beq
\label{6}
\frac{\partial c}{\partial t}=D(t) \nabla^2 c,
\eeq
where $c=n_0/n$ is the concentration of the intruder particles.

In contrast to the usual diffusion equation for molecular (elastic) gases, Eq.\ \eqref{6} cannot be directly integrated in time because of the time dependence of the diffusion coefficient $D$. In the hydrodynamic regime (that is, for times much longer than the mean free time), $D(t)$ depends on time only through its dependence on the granular temperature $T(t)$ \cite{BP04,G19,DG01}. In addition, kinetic theory calculations show that $D(t)$ can be expressed as follows \cite{G19}:
\beq
\label{7}
D(t)=D_0(t) D^*.
\eeq
The (dimensionless) diffusion transport coefficient $D^*$ depends on the parameter space of the system given by the mass ratio $m_0/m$, the diameter ratio $\sigma_0/\sigma$, and the coefficients of restitution $\al$ and $\al_0$. The second quantity on the right hand side,
\beq
\label{8}
D_0(t)=\frac{T(t)}{m_0 \nu(t)},
\eeq
is proportional to the diffusion coefficient for molecular gases. In Eq.\ \eqref{8}, we have introduced the average frequency of collisions between grains (see Appendix \ref{appendixA})
\beq
\label{8.1}
\nu(t)=\frac{\sqrt{2}\pi^{(d-1)/2}}{\Gamma\left(\frac{d}{2}\right)}n\sigma^{d-1}\chi v_\text{th}(t),
\eeq
where $\chi$ is the pair correlation function for grain-grain collisions at contact and $v_\text{th}(t)=\sqrt{2T(t)/m}$ is the thermal velocity of gas particles. An explicit, albeit approximate, expression for the reduced diffusion coefficient $D^*$ can be obtained by considering for instance the first Sonine approximation to the Chapman--Enskog solution \cite{CC70} (this approximation consists in retaining only the leading term in a Sonine polynomial expansion of $f_0$). Its explicit form will be provided in Sec. \ref{sec2}.

Equations \eqref{8} and \eqref{8.1} show that $D_0(t)\propto \sqrt{T(t)}$. As discussed in several previous works \cite{BRGD99,BRCG00,GM04}, the time dependence of the diffusion equation \eqref{6} can be eliminated by introducing a set of appropriate dimensionless time and space variables. One such set is given by
\beq
\label{9}
s=\int_0^t dt' \nu(t'), \quad \mathbf{r}'=\frac{\mathbf{r}}{\ell}.
\eeq
Clearly, the dimensionless time variable $s$ is the number of collisions per gas particle in the time interval between 0 and $t$.  An explicit formula for $s(t)$ is readily obtained by making use of Haff's law \eqref{2} in the expression  $\eqref{8.1}$ of $\nu(t)$ in terms of $v_{th}(t)$. The time integral defining $s(t)$ then gives
\beq
\label{st}
s(t)=2\frac{\nu(0)}{\zeta(0)} \ln \left(1+\frac{1}{2}\,\frac{\zeta(0)}{\nu(0)} t^*\right),
\eeq
where $\nu(0)=(\sqrt{2}\pi^{(d-1)/2}/\Gamma(d/2))n\sigma^{d-1}\chi v_\text{th}(0)$ and  $t^*=\nu(0) t$ is the time in units of the initial intercollisional time $\nu(0)^{-1}$ of the gas particles. The unit length
\beq
\label{9.1}
\ell=\frac{\overline{v}(t)}{\nu (t)}=\frac{\Gamma\left(\frac{d+1}{2}\right)}{\sqrt{2}\pi^{(d-1)/2}}\frac{1}{n\sigma^{d-1}\chi}
\eeq
is the mean free path in a molecular gas of hard spheres \cite{CC70}. Here, $\overline{v}$ is the average speed modulus of a molecular gas at equilibrium, i.e.,
\beq
\label{9.2}
\overline{v}(t)=\frac{\Gamma\left(\frac{d+1}{2}\right)}{\Gamma\left(\frac{d}{2}\right)}v_\text{th}(t).
\eeq
For a three-dimensional gas ($d=3$), one gets the well-known result for hard spheres $\ell=1/(\sqrt{2}\pi \chi n\sigma^2)$.

In terms of the variables $s$ and $\mathbf{r}'$, the diffusion equation \eqref{6} becomes
\beq
\label{10}
\frac{\partial c}{\partial s}=\widetilde{D}\nabla_{\mathbf{r}'}^2 c,
\eeq
where $\nabla_{\mathbf{r}'}^2$ is the Laplace operator in the $\mathbf{r}'$ coordinate and $\widetilde{D}$ is the dimensionless diffusion coefficient
\beq
\label{11}
\widetilde{D}=
\frac{1}{2}\Bigg[\frac{\Gamma\left(\frac{d}{2}\right)}{\Gamma\left(\frac{d+1}{2}\right)}\Bigg]^2 \widetilde D^*,
\eeq
where
\beq
\label{11.1}
\widetilde D^* = \frac{m}{m_0} D^*.
\eeq
Equation \eqref{10} is thus a standard diffusion equation with a \emph{time-independent} diffusion coefficient $\widetilde{D}$. It follows that the MSD of the \emph{intruder's position} $r'$ after a \emph{time} interval $s$ is (see also the end of this section) \cite{L89}
\beq
\label{12}
\langle |\Delta \mathbf{r}'|^2(s)\rangle=2 d \widetilde{D} s,
\eeq
with $\Delta \mathbf{r}'\equiv \mathbf{r}'(s)-\mathbf{r}'(0)$.  Then,
\beq
\label{13}
\frac{\partial}{\partial s}\langle |\Delta \mathbf{r}'|^2(s)\rangle=2 d \widetilde{D}.
\eeq
In terms of the original variables  $\mathbf{r}$ and $t$ [see Eq.\ \eqref{9}], one has
\beq
\label{13.1}
\frac{\partial}{\partial t}\langle |\Delta \mathbf{r}|^2(t)\rangle=2 d \frac{T(t)}{m_0\nu(t)} D^*.
\eeq
Equation \eqref{13} can be seen as a generalization of the Einstein formula relating the diffusion coefficient to the MSD. From Eqs. \eqref{st}, \eqref{11}, and \eqref{12}, one finds that
the expression of $\langle |\Delta \mathbf{r}|^2(t)\rangle$ can be written in terms of the unit length $\ell$ as
\beqa
\label{14}
\langle |\Delta \mathbf{r}|^2(t)\rangle&=& d \Bigg[\frac{\Gamma\left(\frac{d}{2}\right)}{\Gamma\left(\frac{d+1}{2}\right)}\Bigg]^2
\frac{2\nu(0)}{\zeta(0)}\widetilde D^*\nonumber\\
& & \times  \ln \left(1+ \frac{\zeta(0)}{2\nu(0)}t^*\right) \ell^2.
\eeqa
Under the assumptions made (hydrodynamic solution restricted to first-order in $\nabla n_0$), this equation is exact and very general, but $D^*$ and $\zeta(0)$ need to be explicitly determined from the specific parameters of the system at hand; in this context, simplified explicit expressions can be obtained by means of approximations, typically one based on a Sonine polynomial expansion.

In the elastic limit ($\al\to 1$), $\zeta(0)\propto 1-\al^2\to 0$, $\ln \left(1+\zeta(0)t^*/2\nu(0)\right) \approx {\zeta(0)}t^*/{2\nu(0)}$, and so,
\beq
\label{15.1}
\langle |\Delta \mathbf{r}|^2(t)\rangle=d \Bigg[\frac{\Gamma\left(\frac{d}{2}\right)}{\Gamma\left(\frac{d+1}{2}\right)}\Bigg]^2 \widetilde D^*
\ell^2\; t^*.
\eeq
This is the expected result for molecular gases (\emph{normal} diffusion where the MSD is a linear function of time).

For inelastic collisions ($\al\neq 1$ and $\al_0\neq 1$), Eq.\ \eqref{14} shows that the MSD increases \emph{logarithmically} with time, and hence the diffusion turns out to be \emph{ultraslow}. In other words, it is even slower than the typical case of subdiffusion with a MSD $\langle |\Delta \mathbf{r}|^2(t)\rangle \propto t^\beta$ with $0<\beta<1$ \cite{MJChB14}. Equation \eqref{14} applies for arbitrary values of the masses ($m$ and $m_0$) and particle diameters ($\sigma$ and $\sigma_0$), the coefficients of normal restitution ($\al$ and $\al_0$), and the dimensionality $d$ of the system (the dimensionless coefficient $\widetilde D^*$ depends on all these parameters). Note, however, that the time-dependent argument of the logarithm only involves quantities associated with the granular gas and not mechanical properties of the intruder. The reason for this is clear: the time dependence of the MSD is exclusively dictated by $s(t)$ [cf, Eqs.~\eqref{st} and \eqref{12}];  as shown by Eqs.~\eqref{9}, \eqref{8.1}, this quantity is directly obtained from the Haff's cooling law, which only depends on gas properties via the initial cooling rate $\zeta(0)$ and is not influenced by the intruder's mechanical properties.

In the self-diffusion limit case (namely, when $m=m_0$, $\sigma=\sigma_0$, and $\al=\al_0$), the intruder becomes a gas particle like any other, and
Eq.\ \eqref{15.1} agrees with the MSD expression derived by Bodrova \emph{et al.} \cite{BChM15a,BChM15b} for a granular gas with constant coefficient of restitution $\al$. More recently, Blumenfeld \cite{B21} has also derived a logarithmic time dependence of the MSD with the help of a simple random walk model. However, the parameters appearing in Eq.\ \eqref{14} in the self-diffusion case differ from those reported in Ref.\ \cite{B21}. A comparison between both results will be carried out in section \ref{sec3}.

It is worth noting that, from Eq.\ \eqref{10}, it is not only possible to obtain the second-order moment
of $|\Delta \mathbf{r}|$ [see Eq.\ \eqref{12}], but also any other moment of arbitrary order. The solution of Eq.\ \eqref{10} for the delta-peaked initial condition $c(\mathbf{r}',0)=\delta(\mathbf{r}'-\mathbf{r}(0))$ is the $d$-dimensional Gaussian distribution
\beq
\label{15.2}
c(\mathbf{r}',s)=\frac{1}{(4\pi \widetilde{D} s)^{d/2}} e^{-|\mathbf{r}'-\mathbf{r}'(0)|^2/
(4 \widetilde{D} s)}.
\eeq
From this expression one finds
\beqa
\label{15.3}
\langle |\Delta \mathbf{r}'|^{2k}(s)\rangle&=&\int d\mathbf{r}' \, |\Delta \mathbf{r}'|^{2k}
c(\mathbf{r}',s) \nonumber\\
&=&\frac{\Gamma\left(k+\frac{d}{2}\right)}{\Gamma\left(\frac{d}{2}\right)}\,(4\widetilde{D}s)^k,
\eeqa
with $k=1,2,\ldots$. When $k=1$, Eq.\ \eqref{15.3} leads to Eq.\ \eqref{14}.

Before closing this section, it is also convenient to introduce the average intruder-grain collision frequency $\nu_0$. It is given by (see Appendix \ref{appendixA})
\beq
\label{16.0}
\nu_0(t)=\Upsilon \,\nu(t),
\eeq
where
\beq
\label{Upsilon-def}
\Upsilon=\left(\frac{\overline{\sigma}}{\sigma}\right)^{d-1}\frac{\chi_0}{\chi}\left(\frac{1+\theta}{2\theta}\right)^{1/2}.
\eeq
Here, we have defined $\overline{\sigma}=(\sigma+\sigma_0)/2$; the symbol  $\chi_0$ denotes the pair correlation function for intruder-grain collisions at contact, and
\beq
\label{theta}
\theta\equiv \frac{m_0 T}{mT_0}
\eeq
is the ratio between the respective mean square velocities of intruders and grains. By analogy with Eq.\ \eqref{9.1}, the intruder's mean free path is defined as
\beq
\label{16.1}
\ell_0=\frac{\overline{v}_0(t)}{\nu_0 (t)},
\eeq
where $\overline{v}_0$ denotes the intruder's average speed modulus. To compute this average, one has to take into account that the intruder's mean kinetic energy is different from its counterpart for gas particles. This means that $T_0\neq T$, where the partial temperature $T_0$ is a measure of the intruder's mean energy. It is defined as
\beq
\label{18.1}
T_0(t)=\frac{1}{d n_0}\int d\mathbf{v}\; m_0 \mbox{v}^2 f_0(\mathbf{v};t).
\eeq
In principle, the evaluation of the quantity
\beq
\label{v0}
\overline{v}_0=\frac{1}{n_0}\int d\mathbf{v}_0\;  v_0\; f_0(\mathbf{v}_0;t)
\eeq
requires access to the unknown tracer distribution $f_0(\mathbf{v}_0;t)$. In practice, a good estimate
 for $\overline{v}_0$ can be obtained by using the Maxwellian approximation
\beq
\label{16.2}
f_{0,\text{M}}(\mathbf{v}_0;t)=n_0 \left(\frac{m_0}{2\pi T_0(t)}\right)^{d/2}\exp\left(-\frac{m_0v_0^2}{2T_0(t)}\right)
\eeq
for the computation of the integral on the right hand side of Eq.\ \eqref{v0}. One is then left with
\beq
\label{16.3}
\overline{v}_0(t)=\frac{\Gamma\left(\frac{d+1}{2}\right)}{\Gamma\left(\frac{d}{2}\right)} \theta^{-1/2}v_\text{th}(t)
=
\frac{\overline{v}(t)}{\sqrt{\theta}}.
\eeq
Finally, substitution of Eqs.\ \eqref{16.0} and \eqref{16.3} into Eq.\ \eqref{16.1} leads to a relationship between the mean free paths of the two particle species:
\beq
\label{16.4}
\ell_0=\frac{\ell}{\Upsilon \sqrt{\theta}}.
\eeq


\section{First-Sonine approximation to $\widetilde D^*$}
\label{sec2}

In order to obtain the explicit dependence of the MSD on the space parameters of the problem, one still needs an explicit expression for
the (dimensionless) diffusion coefficient $\widetilde D^*$. In a way similar as for elastic collisions \cite{CC70}, the form of $D^*$ is given in terms of the solution of a linear integral equation which can be approximately solved by considering the first few terms of a Sonine polynomial expansion. Usually, only the leading term in this Sonine expansion (the so-called first Sonine approximation) is retained to estimate $\widetilde D^*$. In spite of the simplicity of this approach, it generally agrees excellently with Monte Carlo simulations, even for moderately strong inelasticities (see for instance, Figs.\ (6.2)--(6.5) of the textbook \cite{G19}).

The expression of the reduced coefficient $\widetilde D^*$  in the first-Sonine approximation can be written as \cite{GM04,GM07}
\beq
\label{16}
\widetilde D^*=\frac{1}{\theta(\nu_D^*-\frac{1}{2}\zeta^*)},
\eeq
where
\beq
\label{17}
\nu_D^*=\frac{ 2 }{d}  \mu (1+\al_0) \Upsilon,
\eeq
\beq
\label{18}
\zeta^*=\frac{\zeta(t)}{\nu(t)}=\frac{1-\al^2}{d}.
\eeq
Here, $\mu=m/(m+m_0)$. For dilute granular gases ($n\sigma^d\to 0$), $\chi=\chi_0=1$.

In general, the temperature ratio $T_0/T$ is different from unit (breakdown of energy equipartition). It is obtained from
the condition
\beq
\label{20.1}
\zeta_0^*=\zeta^*,
\eeq
where $\zeta^*$ is given by Eq.\ \eqref{18}, and the ``partial'' (dimensionless) cooling rate $\zeta_0^*$ characterizing the rate of energy dissipation for intruder-grain collisions is \cite{G19}
\beqa
\label{21}
\zeta_0^*&=&\frac{2\sqrt{2}}{d} \mu \frac{\chi_0}{\chi} \left(\frac{\overline{\sigma}}{\sigma}\right)^{d-1} \left(\frac{1+\theta}{\theta}\right)^{1/2}(1+\al_{0})
\nonumber\\
& & \times
\left[1-\frac{1}{2}\mu (1+\theta)(1+\al_{0})\right].
\eeqa
Insertion of Eqs. \eqref{18} and \eqref{21} into Eq. \eqref{20.1} leads to a cubic equation for $\theta$ with three different solutions. Among these, we choose the one allowing us to recover the correct value of $\theta$ for elastic collisions (i.e., $\theta=m_0/m$ if $\alpha=\alpha_0=1$). When intruder and grains are mechanically equivalent ($m=m_0$, $\sigma=\sigma_0$, $\al=\al_0$), the requirement $\zeta_0^*=\zeta^*$ yields $T=T_0$ (energy equipartition). Beyond the self-diffusion limit case, the inequality $T\neq T_0$ resulting from the breakdown of energy equipartition turns out to have a significant impact on the properties of diffusive transport \cite{G19,GM04,GM07}.

Equation \eqref{18} tell us that, in the first Sonine approximation, $\zeta(0)/\nu(0)=(1-\al^2)/d$.  We can use this result to obtain explicit expressions for the number of collisions per gas particle (cf.~Eq.~\eqref{st})
 \beq
\label{stexp}
s(t)=\frac{2d}{1-\alpha^2}\ln \left(\!1+\frac{1-\alpha^2}{2d}t^*\!\right),
\eeq
and the scaled intruder's MSD (cf.~Eq.~\eqref{14})
\beq
\label{21.0}
\frac{\langle |\Delta \mathbf{r}|^2(t)\rangle}{\ell^2}=2 \Bigg[\frac{d\Gamma\left(\frac{d}{2}\right)}{\Gamma\left(\frac{d+1}{2}\right)}\Bigg]^2
\frac{\ln \left(1+\frac{1-\al^2}{2d} t^*\right)}{\theta\left(\nu_D^*-\frac{1-\al^2}{2d}\right)\left(1-\al^2\right)},
\eeq

\subsection{Elastic collisions}

For elastic collisions ($\al=\al_0=1$), $\zeta^*=0$, $T_0=T$, one obtains the following diffusion coefficient:
\beq
\label{21.1}
D=\frac{d\Gamma\left(\frac{d}{2}\right)}{4\sqrt{2}\pi^{(d-1)/2}}\frac{1}{n\overline{\sigma}^{d-1}\chi_0}\sqrt{\frac{(m+m_0)T}{m m_0}}.
\eeq
The expression \eqref{21.1} of $D$ agrees with the known result for molecular mixtures of hard spheres \cite{CC70}.

\subsection{Self-diffusion}   

As already mentioned, in the self-diffusion limit case ($m=m_0$, $\sigma=\sigma_0$, $\chi=\chi_0$, and $\al=\al_0$), $T_0=T$ and Eq.\ \eqref{16} reads
\beq
\label{21.2}
\widetilde D^*=\frac{2d}{(1+\al)^2}.
\eeq
Hence, the self-diffusion coefficient $D$ is
\beq
\label{21.2.1}
D(t)=\frac{d \Gamma\left(\frac{d}{2}\right)}{\pi^{(d-1)/2}}\frac{1}{(1+\al)^2}\frac{1}{n\sigma^{d-1}\chi}\sqrt{\frac{T}{m}}.
\eeq
This expression agrees with the one obtained in Refs.\ \cite{BRCG00} and \cite{BB12}.

\subsection{Brownian diffusion}
\vicente{The Brownian diffusive regime is characterized by the conditions $m_0\gg m$ and $T_0/T$
finite. In this case}, the partial cooling rate $\zeta_0=\zeta_0^* \nu$ can be written as
\beqa
\label{21.3}
\zeta_0&=&  2 a \left(1-a\frac{T}{T_0}\right)\gamma_e,
\eeqa
where
\beq
\label{21.4}
a=\frac{1+\al_0}{2}, \quad
\gamma_e(t)=\frac{4\pi^{(d-1)/2}}{d\Gamma\left(\frac{d}{2}\right)}\chi_0 \frac{m}{m_0} n\overline{\sigma}^{d-1}\sqrt{\frac{2T(t)}{m}}.
\eeq
Taking Eq.\ \eqref{21.3} into account, the condition $\zeta=\zeta_0$ for determining the temperature ratio $T_0/T$ gives
\beq
\label{21.5}
\zeta=2 a \left(1-a\frac{T}{T_0}\right)\gamma_e,
\eeq
whose solution is
\beq
\label{21.6}
\frac{T_0}{T}=\frac{a}{1-\lambda_e}, \quad \lambda_e=\frac{\zeta}{2a\gamma_e}.
\eeq
The expression \eqref{21.6} agrees with the results derived by Brey \emph{et al.} in their study of Brownian motion in a granular gas \cite{BRGD99}.

In addition, it is straightforward to prove that in the Brownian limit the scaled diffusion coefficient is
\beq
\label{21.7}
D^*=\frac{\nu(t)}{\gamma_e(t)}\frac{1}{(1-\lambda_e)^2}.
\eeq
To derive Eq.\ \eqref{21.7}, we have taken into account Eq.\ \eqref{21.6}. Thus, the diffusion coefficient $D$ in the Brownian limit is
\beq
\label{21.8}
D(t)=\frac{T(t)}{m_0 \gamma_e(t)}\frac{1}{(1-\lambda_e)^2}.
\eeq
Equation \eqref{21.8} is consistent with the results obtained in Ref.\ \cite{BRGD99}.

\vicente{It is worth mentioning that there is a ``nonequilibrium" phase transition \cite{SD01,SD01a}
in the limit of a very massive intruder, $m_0/m\to \infty$, which corresponds to an extreme violation
of energy equipartition. More precisely, there is a region in parameter space
 where $\theta^{-1}\equiv m_0T/(m T_0)$ tends to infinity as $m_0/m\to\infty$,
 while $T_0/T$ remains finite \cite{SD01a}.  In this Brownian diffusive regime, the diffusion coefficient remains finite. However, there is another region where the limit $m_0/m\to\infty$ implies the divergence of $T_0/T$ \cite{SD01a} and formally yields an infinite value of the diffusion coefficient. This behaviour reflects the onset of a ballistic regime and the breakdown of the diffusive description arising from the Boltzmann-Lorentz framework.
 }

\vicente{A formal description of the Brownian diffusive regime is provided by a Fokker-Planck equation
\cite{BDS99} and the corresponding Langevin equation (Brownian motion with inertia)\cite{K61,Sarracino10}. The effective damping coefficient in this Langevin equation
[written in terms of the dimensionless time $s(t)$] turns out to be equal to $a\gamma_e/\nu$, and is directly related to the inverse of the characteristic decay time of the intruder's velocity autocorrelation function.}

\section{Intruder's motion as a random walk}
\label{Sec-RW}

Until now, the treatment of intruder diffusion has been based on the Chapman-Enskog solution to the Boltzmann kinetic equation. This is the most usual (and successful) way of dealing with the diffusion problem in molecular and granular gases. Though much less common and successful, there is an alternative approach to study molecular gases: free-path theory. In this approach, the motion of gas particles is regarded as a succession of random flights between collisions, and one treats the diffusive problem as a random walk problem. This approach provides a simple and intuitive physical picture of transport phenomena in gases \cite{Jeans1959,Furry1951,Reif1965}, hence its appeal. On the other hand, when correctly applied to molecular gases, free-path theory yields results identical to those obtained from the Chapman-Enskog solution \cite{Furry1951,Yang1949,Monchick1962}.

At a microscopic level, the intruder's motion can be seen as a succession of ballistic displacements $\Bell_i$ interrupted by collisions with the gas particles. This type of motion belongs to the class of random walks called random flights: the intruder is identified with a random walker that moves ballistically and changes its direction randomly every time a collision takes place. The intruder's displacement after $N$ collisions is
\beq
\label{RW.1}
\Delta \mathbf{r}=\sum_{i=1}^N \Bell_i.
\eeq
Correspondingly, the MSD of the random walker reads
\beq
\label{MSDlesA}
\langle |\Delta \mathbf{r}(N)|^2\rangle = N \langle \ell^2 \rangle   +\sum_{i\neq j}^N \langle \Bell_i\cdot \Bell_j\rangle,
\eeq
where $\langle \ell^2 \rangle\equiv \langle \ell_1^2 \rangle=\langle \ell_2^2 \rangle=\cdots$.

If one assumes that the changes of direction are drawn from an \emph{isotropic} distribution, then $\langle \Bell_i\cdot \Bell_j\rangle=0$; the intruder's MSD can then be approximated by $ \langle |\Delta \mathbf{r}(N)|^2\rangle = N \langle \ell^2 \rangle$ or, in terms of time, by the expression
\beq
\label{MSD0}
\langle |\Delta \mathbf{r}|^2(t)\rangle= s_0(t) \langle \ell^2 \rangle.
\eeq
Here, $s_0(t)$ denotes the average number of steps of the random walker (or, equivalently the average number of intruder's collisions) up to time $t$ \cite{Reif1965}.
In the self-diffusive case ($m=m_0$ and $\sigma=\sigma_0$) of molecular gases ($\alpha=1$), Eq.~\eqref{stexp} yields $s_0(t)=s(t)=t^*\equiv \nu(0)t=t \overline{v}/\ell$, and so $\langle \ell^2 \rangle=2\ell^2$, according to the
elementary kinetic theory of gases \cite{McQuarrie1975}. From this relation, one finds
\beq
\label{RW.2}
\langle |\Delta \mathbf{r}|^2(t)\rangle=2\overline{v}\ell t.
\eeq
This is in fact the expression provided by the elementary kinetic theory of gases for the MSD \cite{Reif1965}. On the other hand, for elastic collisions, Eq.\ \eqref{13.1} yields
\beq
\label{RW.3}
\langle |\Delta \mathbf{r}|^2(t)\rangle=2d D t,
\eeq
where use has been made of the fact that the intruder and the gas particles are mechanically equivalent. From Eqs.\ \eqref{RW.2} and \eqref{RW.3}, one can identify the form of the self-diffusion coefficient $D$:
\beq
\label{RW.4}
D=\frac{1}{d}\overline{v}\ell.
\eeq
However, it is well known that the result \eqref{RW.4} is poor: it is smaller than the value of the first-Sonine approximation [see Eq.\ \eqref{21.2.1} when $\al=1$] by a factor of $16/9\pi\approx 0.566$ for $d=3$.

The reason for the above discrepancy is clear: Equation \eqref{MSD0} was obtained by assuming that the intruder's deflections due to collisions are isotropically distributed, so that $\langle \Bell_i\cdot \Bell_j\rangle=0$. Of course, this assumption is in general not true, since there is a correlation between the pre- and the post-collisional directions of the particle velocities. This necessarily implies $\langle \Bell_j\cdot \Bell_{j+1}\rangle\neq 0$. For instance, in the self-diffusion problem, when the collisions are elastic the average fraction of the initial velocity that survives after a collision (a quantity known as \emph{persistence}) is about 0.406 for $d=3$ \cite{CC70}. In other words, forward collisions are more likely than backward ones, and consequently $\langle \Bell_j\cdot \Bell_{j+1}\rangle> 0$.

However, after a few collisions, the intruder's velocity (or equivalently, its displacement $\Bell_i$) becomes uncorrelated with the initial one, and so the overall random motion \emph{is} isotropic on sufficiently large length and time scales. Thus, after a short initial regime, the particle's motion is well described by the isotropic diffusion equation \eqref{6}.


Including the correlation terms $\langle \Bell_i\cdot \Bell_{j}\rangle$, Eq.\ \eqref{MSDlesA} can be rewritten as
\beq
\label{MSDles3}
 \langle |\Delta \mathbf{r}(N)|^2\rangle = N \ell_e^2,
\eeq
where
\beq
\label{ell2Sum}
\ell_e^2=\langle \ell^2\rangle +\frac{1}{N}\sum_{i\neq j}^N \langle \Bell_i \cdot \Bell_j \rangle.
\eeq
In terms of time (or the number of steps $s_0(t)$), Eq.\ \eqref{MSDles3} reads
\beq
\label{MSDles}
\langle |\Delta \mathbf{r}|^2(t)\rangle= s_0(t)\, \ell_e^2.
\eeq
Interestingly, the MSD obtained from this diffusion equation can be viewed as resulting from an \emph{equivalent} random walk with isotropic steps of size $\ell_e$. This fictitious random walk yields \emph{the same MSD} as the actual random walk, which is characterized by strongly anisotropic changes in direction as a result of the collision rules for smooth hard spheres. However, the effective length $\ell_e$ of the displacements between consecutive steps of the isotropic walk differs from the intruder's mean free path $\ell_0$ owing to the persistence induced by the microscopic collision rule (forward collisions are more likely than backward ones). In this sense, $\ell_e$ is a measure of the persistence of the actual random walk.

An evaluation of  the effective step length $\ell_e $ based on  microscopic  arguments is not at all easy, even for the simplest case of self-diffusion with elastic collisions \cite{Furry1951,Yang1949}.  However, we can get $\ell_e$ as a byproduct of Eq.\ \eqref{21.0}.  Indeed, from Eqs. \eqref{st} and \eqref{16.0}, we find
\beqa
\label{s0self}
s_0(t)&=&\int_0^t \nu_0(t') dt'\nonumber\\
&=& \Upsilon s(t)=
\frac{2\Upsilon d}{1-\alpha^2}\ln \left(\!1+\frac{1-\alpha^2}{2d}t^*\!\right),
\eeqa
where use has been made of the result $\zeta(0)/\nu(0)=(1-\al^2)/d$. The expression of $\ell_e^2$ can be easily identified by inserting Eq.\ \eqref{s0self} into the right hand side of Eq.\ \eqref{MSDles} and comparing the resulting equation with Eq.\ \eqref{21.0}.
One is then left with
\beqa
\ell_e^2&=&d\Bigg[\frac{\Gamma\left(\frac{d}{2}\right)}{\Gamma\left(\frac{d+1}{2}\right)}\Bigg]^2
\left(\frac{\sigma}{\overline{\sigma}}\right)^{d-1}\frac{\chi}{\chi_0}\left(\frac{2\theta}{1+\theta}\right)^{1/2}  \nonumber \\
&&\times  \frac{\ell^2}{\theta\left(\nu_D^*-\frac{1}{2}\zeta^*\right)},
\eeqa
where the explicit form of $\Upsilon$ [see Eq.\ \eqref{Upsilon-def}] has been used.

In the particular case of self-diffusion, simplified expressions are obtained, namely,  $s_0(t)=s(t)$ and
\beq
\label{ellself}
\ell_e=
\frac{d\Gamma\left(\frac{d}{2}\right)}{\Gamma\left(\frac{d+1}{2}\right)}\,\frac{\sqrt{2}}{1+\alpha}\,
\ell.
\eeq
\vicente{In the Brownian diffusive regime ($m\gg m_0$ and $T_0/T$ finite), $\ell_e$ reads
\beq
\label{ellbrow}
\ell_e=
\frac{d\Gamma\left(\frac{d}{2}\right)}{\Gamma\left(\frac{d+1}{2}\right)}
\frac{\sqrt{2}}{1+\alpha_0}\,
\left(\frac{\sigma}{\overline{\sigma}}\right)^{d-1}\,\frac{\chi }{\chi_0 }\,\frac{T_0}{ T}\,
\ell.
\eeq
Since the temperature ratio remains finite in the diffusive regime, so does $\ell_e$ too.
In contrast, in the region where $T_0/T\to\infty$, the effective length $\ell_e$ tends to infinity
as expected.}

Beyond the above two limiting cases, the dependence of $\ell_e^2$ and $s_0$ on the parameter space will prove useful for the physical discussion in Secs. \ref{sec3} and \ref{sec4}.

\section{Self-diffusion problem. Comparison with Blumenfeld's results}
\label{sec3}

\begin{figure}
\begin{center}
\includegraphics[width=.7\columnwidth]{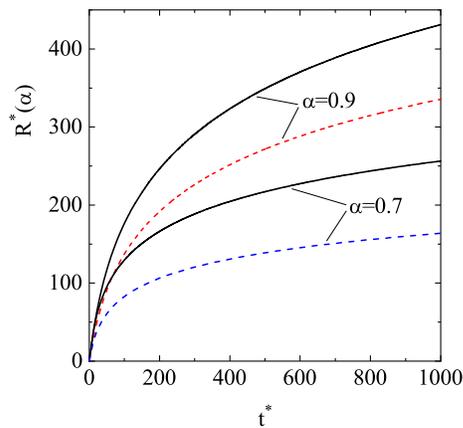}
\end{center}
\caption{MSD scaled with the squared mean free path $R^*\equiv \langle |\Delta \mathbf{r}|^2\rangle/\ell^2$ versus the (reduced) time $t^*\equiv t/t_0$ for two different values of the coefficient of normal restitution $\al$. The solid lines refer to the results derived here [Eq.\ \eqref{24}], while the dashed lines correspond to the results obtained by Blumenfeld [Eq.\ \eqref{28}]. Intruders and grains are assumed to be mechanically equivalent.
\label{fig1}}
\end{figure}
\begin{figure}
\begin{center}
\includegraphics[width=.7\columnwidth]{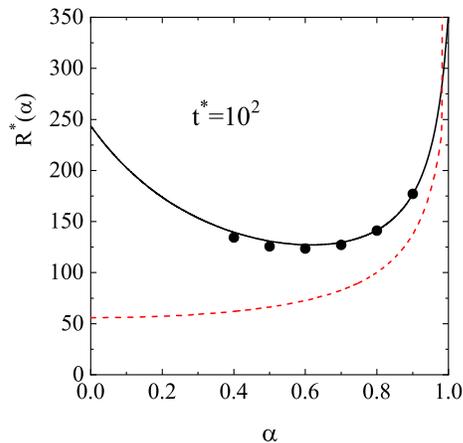}
\end{center}
\caption{Plot of the MSD scaled with the squared mean free path $R^*\equiv \langle |\Delta \mathbf{r}|^2\rangle/\ell^2$ as a function of the coefficient of normal restitution $\al$ for $t^*=10^2$ when intruders and grains are assumed to be mechanically equivalent. The solid line is the result obtained from Eq.\ \eqref{24}, while the dashed line represents the result \eqref{28} obtained from Blumenfeld's approach. The symbols refer to the results obtained from Monte Carlo simulations.
\label{fig2}}
\end{figure}

Before considering the general case, we study in this section the MSD when the intruder and gas particles are mechanically equivalent (self-diffusion case). In particular, we seek to obtain a close comparison with recent results derived by Blumenfeld \cite{B21} by means of a random walk model. To this end, we will consider the system studied by Blumenfeld \cite{B21}, i.e., a dilute ($\chi=\chi_0=1$) granular gas of hard spheres ($d=3$). In the self-diffusion case ($m=m_0$, $\sigma=\sigma_0$, and $\al=\al_0$), $T_0=T$, and so $\theta=1$. Thus, Eq.\ \eqref{21.0} yields
\beq
\label{24}
\frac{\langle |\Delta \mathbf{r}|^2(t)\rangle}{\ell^2}=\frac{27 \pi}{(1-\al)(1+\al)^3}\ln \Bigg[1+\frac{1-\al^2}{6} t^*\Bigg].
\eeq
Note that this result also follows by inserting Eqs. \eqref{ellself} and \eqref{s0self} into Eq. \eqref{MSDles}, then taking into account that $\Upsilon=1$ for self-diffusion, and eventually setting $d=3$.


To make a clean comparison with Blumenfeld's results for the MSD \cite{B21}, one first has to note that he performs a mean field analysis of self-diffusion in a gas of identical particles subject to hard-core interactions. Blumenfeld's random walk model uses the mean free path $l_0\approx 0.55396 n^{-1/3}$, which corresponds to the mean distance between $nV$ points that are randomly scattered  in a volume $V$. For a fixed value of the initial velocity modulus $v_0$, Blumenfeld's expression for the MSD is as follows:
\beq
\label{RaphaelMSD}
\frac{\langle |\Delta \mathbf{r}|^2 \rangle}{l_0^2}= \frac{3}{\ln(1/\epsilon)}\ln(1+\lambda v_0),
\eeq
where
\beq
\lambda\equiv \lambda(\epsilon,t)=\frac{1-\epsilon}{\epsilon \, l_0}\,t,
\eeq
and $\epsilon$ denotes the (constant) ratio between the pre-collisional and the post-collisional velocity moduli. \vicente{Note that $\epsilon$ and $\alpha$ are different quantities, since the latter refers to the ratio of the \emph{normal components} of the velocities}

In order to compare Eq.\ \eqref{RaphaelMSD} with Eq. \ \eqref{24}, we must first evaluate the average $\langle \ln(1+\lambda v_0) \rangle $ over the Maxwellian distribution $f_\text{M}(\mathbf{v}_0)$:
\beq
\label{M.1}
\langle \ln(1+\lambda v_0) \rangle=\frac{1}{n_0}\int d\mathbf{v}_0\; \ln(1+\lambda v_0)\; f_\text{M}(\mathbf{v}_0),
\eeq
where
\beq
\label{M.2}
f_\text{M}(\mathbf{v}_0)=n_0\left(\frac{m}{2\pi T(0)}\right)^{3/2} \exp\left(-\frac{m v_0^2}{2T(0)}\right),
\eeq
$T(0)$ being the initial temperature. Some technical details regarding the computation of the average \eqref{M.1} are provided in Appendix \ref{appendixB}.
As it turns out, the average of $\ln(1+\lambda v_0)$ over such a distribution is very well approximated from above by $\ln(1+\lambda \overline{v}_0)$ [the initial average velocity $\overline{v}_0\equiv\overline{v}(0)=\sqrt{8 T(0)/(\pi m)}$ is obtained by setting $d=3$ in Eq.\ \eqref{9.2}]. In fact, the relative deviation of these two quantities as a function of $\lambda$ never exceeds $4\%$,  and one can show that $\langle \ln(1+\lambda v_0) \rangle\to \ln(1+\lambda \overline{v}_0)$ as $\lambda\to\infty$ (or, equivalently, as $t\to\infty$ when the rest of parameters are fixed) [see Appendix \ref{appendixB}]. With the aforementioned approximation, Eq. \ \eqref{RaphaelMSD} becomes
\beq
\label{AvRaphaelMSD}
\frac{\langle |\Delta \mathbf{r}|^2 \rangle}{l_0^2}= \frac{3}{\ln(1/\epsilon)}
\ln\left[1-\frac{1-\epsilon}{\epsilon}\,\left(\frac{\overline{v}_0}{l_0}\right) t \right].
\eeq

The next step to achieve a fair comparison between Eq.\ \eqref{24} and Blumenfeld's result is to replace the mean free path $l_0$ in Eq. \ \eqref{AvRaphaelMSD} with the mean free path for a molecular gas of hard spheres $\ell=1/(\sqrt{2}\pi n\sigma^2)$ [cf. Eq.\ \eqref{9.1} with $d=3$]. We hereby aim to account for effects related to the finite size of the spheres within the limitations imposed by a mean field approach.

Furthermore, while the (constant) parameter $\epsilon$ introduced in Blumenfeld's model accounts for the loss of momentum of each colliding particle, the coefficient of normal restitution $\al$ accounts for the reduction of the magnitude of the normal component of the post-collisional relative velocity with respect to its pre-collisional counterpart. To express $\epsilon$ in terms of $\al$, we fix the value of $\epsilon$ by requiring that Blumenfeld's formula for Haff's law coincide with ours [Eq.\ \eqref{2}]. This choice yields $\epsilon=6/(7-\al^2)$. Inserting this expression into Eq. \ \eqref{AvRaphaelMSD} and making the replacement $l_0\to\ell$ in the resulting equation, we find
\beq
\label{28}
\frac{\langle |\Delta \mathbf{r}|^2(t^*)\rangle}{\ell^2}=\frac{3}{\ln \left(\frac{7-\al^2}{6}\right)}\ln\left(1+\frac{1-\al^2}{6} t^*\right),
\eeq
where we have made use of the relation $\nu(0)=\overline{v}_0/\ell$ [cf. Eq. \ \eqref{9.1}] and of the definition $t^*=\nu(0) t$.

We are now in the position to make a fair comparison between Blumenfeld's model and our result \eqref{24}. For elastic collisions ($\al\to 1$), Eq.\ \eqref{28} yields
\beq
\label{29}
\lim_{\al\to 1}\frac{\langle |\Delta \mathbf{r}|^2(t^*)\rangle}{\ell^2}=3\,t^*,
\eeq
while in the elastic limit, Eq.\ \eqref{24} yields
\beq
\label{27.1}
\lim_{\al\to 1}\frac{\langle |\Delta \mathbf{r}|^2(t^*)\rangle}{\ell^2}=\frac{9\pi}{8}\,t^*\approx 3.53\,t^*.
\eeq
Eq.\ \eqref{27.1} differs from Blumenfeld's result \eqref{29}, but is consistent with the known results for molecular gases \cite{CC70}.

Figure \ref{fig1} shows the dependence of the (scaled) MSD $R^*\equiv \langle |\Delta \mathbf{r}|^2\rangle/\ell^2$ on the (reduced) time $t^*$ for two different values of the coefficient of normal restitution $\al$. Solid and dashed lines are obtained from Eqs.\ \eqref{24} and \eqref{28}, respectively. Clearly, there are significant quantitative differences between our results and those based on Blumenfeld's approach \cite{B21}.

To complement Fig.\ \ref{fig1}, Fig.\ \ref{fig2} shows $R^*$ versus $\al$ for \vicente{$t^*=10^{2}$}. Solid and dashed lines refer to the theoretical results obtained form Eqs.\ \eqref{24}  and \eqref{28}, respectively. Symbols correspond to the DSMC results obtained by replacing in the expression \eqref{14} $\widetilde D^*$ with its corresponding simulation value (reported in Ref.\ \cite{BRCG00}). An excellent agreement between the present theory and DSMC simulations is found. It is important to note that the (scaled) diffusion coefficient $D^*$ (recall that $\widetilde D^*=D^*$ in the self-diffusion problem) is measured in the simulations \cite{BRGD99,BRCG00,GM04} through the relationship \eqref{13.1}; the linear relation between the intruder's MSD $\langle|\Delta \mathbf{r}'|^2\rangle$ after a ``time'' interval $s$ confirms the validity of the logarithmic law \eqref{14}. While Blumenfeld's expression for $R^*$ exhibits a monotonic dependence on the inelasticity parameter ($R^*$ decreases with decreasing $\al$), in our case $R^*$ depends non-monotonically on $\al$. Notwithstanding this, it is quite remarkable that Blumenfeld's mean field model is able to capture the correct qualitative time dependence of the MSD (given by a logarithmic law fully consistent with Haff's law) just by making use of a few simple assumptions.

It is interesting to note that the qualitative behaviour depicted by the solid line in
Fig. \ \ref{fig2} only holds as long as $t^*>6$. Otherwise, the minimum is suppressed, and Eq.\ \eqref{24} yields a monotonically decreasing $R^*$ with increasing $\alpha$. For $t^*=6$, the $\alpha$-derivative of $R^*$ vanishes at $\alpha_{min}=1$, whereas at longer times $\alpha_{min}$ is shifted towards smaller values that tend to $\alpha_{min}=1/2$ as $t^*\to \infty$. Thus, regardless of the value of $t^*$,  in the $\alpha$-range [0,0.5] the MSD always decreases with increasing  $\alpha$.

The $R^*(\alpha)$-behaviour displayed in Fig. \ \ref{fig2} can be intuitively understood by noting that there are two competing effects.
On the one hand, since the root-mean-square velocity is proportional to $\sqrt{T(t)}$, one expects that $R^*$ will decrease as the collisions become more inelastic, i.e., when $\al$ decreases. Note that this reduction in the speed of the particles implies a reduction in the number of collisions that occur up to a time $t$, that is, a reduction in $s_0(t)$. The dependence of $s_0(t)$ on $\alpha$ is mainly described by the factor $(1-\alpha^2)^{-1}$  of Eq.~\eqref{stexp}.
On the other hand, when collisions become more inelastic, the average angle between the velocity of a particle before and after the collision is reduced, which makes its trajectory straighter. In other words, as inelasticity increases, the dispersion is less effective (i.e., more persistent), resulting in an increased MSD \cite{BP04}. This effect explains the increase of $\ell_e^2$ as $\alpha$ decreases, in agreement with Eq.~\eqref{ellself} since $\ell_e^2\propto(1+\alpha)^{-2}$. Therefore, depending on which of these two effects prevails, $\langle |\Delta \mathbf{r}|^2\rangle=s_0(t)\ell_e^2$ will increase or decrease with increasing inelasticity. Figure \ref{fig2} shows that the first effect is dominant for a not too strong inelasticity ($\al>\alpha_{min}\approx 0.5$), while the second effect prevails for $\al<\alpha_{min}$ ( the larger $t$ is, the closer $\alpha_{min}$ gets to $1/2$.) .

\section{Diffusion problem}
\label{sec4}

In the general case, the parameter space includes the respective mass and diameter ratios $m_0/m$ and $\sigma_0/\sigma$, the coefficients of normal restitution $\al$ and $\al_0$, the reduced density $n\sigma^d$, and the dimensionality $d$. Because of the many parameters involved,  for our purposes a full presentation of the obtained results would result in an excessive level of detail. As done in Sec. \ref{sec3}, we therefore focus again on the case of a dilute granular gas ($\chi=\chi_0=1$) of hard spheres ($d=3$). In spite of these restrictions, we show below that the observed phenomenology is already quite intricate in this particular case.

\begin{figure}
\begin{center}
\includegraphics[width=.7\columnwidth]{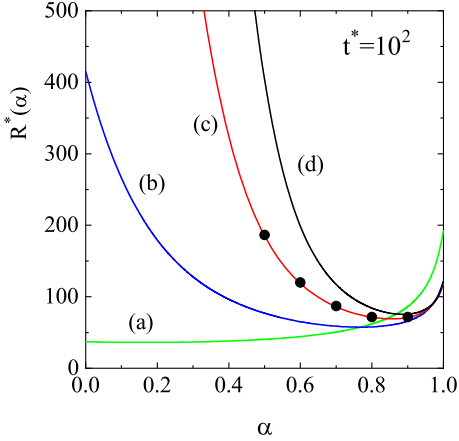}
\end{center}
\caption{Plot of the MSD scaled with the squared mean free path of the 3d granular gas $R^*\equiv \langle |\Delta \mathbf{r}|^2\rangle/\ell^2$ as a function of the coefficient of normal restitution $\alpha=\alpha_0$ for $\sigma_0/\sigma=2$, $t^*=10^2$, and four different values of the mass ratio: $m_0/m=1/2$ (a), $m_0/m=5$ (b), $m_0/m=8$ (c), and $m_0/m=10$ (d). The symbols refer to the results obtained from Monte Carlo simulations for $m_0/m=8$.
\label{fig3}}
\end{figure}

\begin{figure}
\begin{center}
\includegraphics[width=.7\columnwidth]{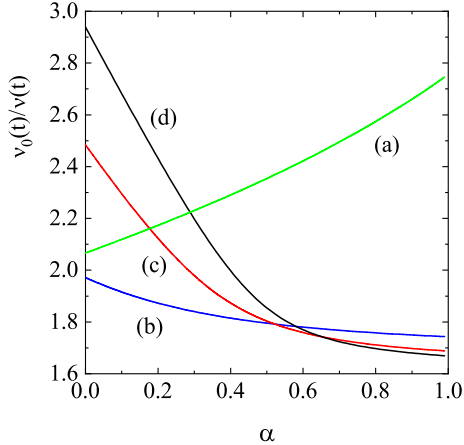}
\end{center}
\caption{Plot of $\Upsilon=\nu_0/\nu$ as a function of the coefficient of normal restitution $\al=\al_0$ for the cases depicted in Fig. \ref{3}, i.e., $m_0/m=1/2$ (a), $m_0/m=5$ (b), $m_0/m=8$ (c), and $m_0/m=10$ (d). In all cases, $\sigma_0/\sigma=2$.
\label{fig4}}
\end{figure}

The dependence of the (scaled) MSD $R^*$ on the (common) coefficient of restitution $\al=\al_0$ is shown in Fig.\ \ref{fig3} for $t^*=10^5$, $\sigma_0/\sigma=2$ and four different values of the mass ratio $m_0/m$. DSMC results obtained in Ref.\ \cite{GM04} for $m_0/m=8$ are also plotted. As in the case of Fig.\ \ref{fig2}, the theory agrees very well with the simulation results, thereby confirming the accuracy of the first Sonine approximation to $D^*$ for this case. For a given value of the mass ratio, $R^*$ again exhibits a non-monotonic dependence on $\al$.

On the other hand, Fig.\ \ref{fig3} also shows that, for any given value of $\al$ (the common coefficient of normal restitution), the scaled MSD always increases with increasing $m_0/m$  as long as  $m_0>m$ (for $m_0<m$ we find that this is no longer the case).

In view of our previous discussion for the self-diffusive case, it seems natural to enquire to what extent this behaviour at fixed $t^*$ is related to the mass dependence of the respective collision frequencies. Figure \ref{fig4} shows the intruder-grain collision frequency ratio $\nu_0/\nu$ as a function of $\al$ for the systems considered in Fig.\ \ref{fig3}. For not too strong inelasticity ($\al \gtrsim 0.6$, say) the frequency ratio $\nu_0/\nu$ increases with decreasing $m_0/m$, but the opposite happens for very inelastic systems. However, in spite of the different dependence of $\nu_0/\nu$ on $m_0/m$ for different $\alpha$, we already know from Fig.\ \ref{fig3} that, for any fixed value of $\al$, the scaled MSD $R^*$ always increases with increasing mass ratio when $m_0/m>1$. Thus, the mass dependence of $\nu_0/\nu$ does not seem to play a major role in the qualitative behaviour observed for small enough $\alpha$ (for $m_0/m>1$, even though the collision frequency increases with increasing mass ratio, the MSD grows).

We ascribe this behaviour to the increasing deviation from energy equipartition (measured by the departure of the temperature ratio $T_0/T$ from unity) with increasing mass ratio $m_0/m$ \cite{GD99}. A larger value of $T_0/T$ implies an augmented persistence of the intruder's motion and the corresponding increase of $R^*$ with growing $m_0/m$. Thus, beyond the impact of the relative collision frequency on $R^*$, the inertia generated by the difference in kinetic energy between the intruder and the gas particles appears to be the dominant effect here.

\begin{figure}
\begin{center}
\includegraphics[width=.7\columnwidth]{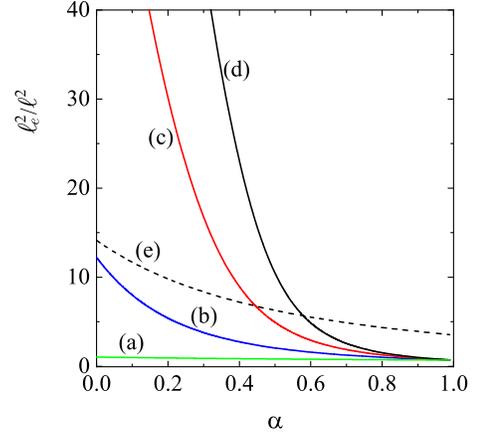}
\end{center}
\caption{Plot of the square effective length $\ell_e^2/\ell^2$  scaled with the mean free path of the gas particles versus the (common) coefficient of normal restitution $\al=\al_0$. The solid lines correspond to the four different cases depicted in Fig. \ref{fig3}, namely, $m_0/m=0.5$ (a), $m_0/m=5$ (b), $m_0/m=8$ (c), and $m_0/m=10$ (d). In all four cases, the diameter ratio is $\sigma_0/\sigma=2$. The dashed line (e) corresponds to the  self-diffusion case with $m_0/m=1$ and $\sigma_0/\sigma=1$.
\label{sqlealpha}}
\end{figure}

\begin{figure}
\begin{center}
\includegraphics[width=.7\columnwidth]{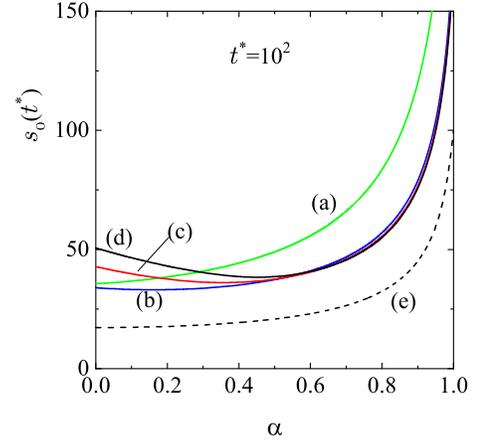}
\end{center}
\caption{Plot of the number of steps $s_0$ versus the (common) coefficient of normal restitution $\al=\al_0$. Solid lines correspond to the four different cases depicted in
Fig. \ref{fig3}, namely, $m_0/m=0.5$ (a), $m_0/m=5$ (b), $m_0/m=8$ (c), and $m_0/m=10$ (d). In all four cases, the diameter ratio is $\sigma_0/\sigma=2$. The dashed line (e) corresponds to the  self-diffusion case with $m_0/m=1$ and $\sigma_0/\sigma=1$. For all curves one has $t^*=10^2$.
\label{s0alpha}}
\end{figure}

Let us now rationalize the behaviour depicted in Fig. \ref{fig3} in terms of the effective step length $\ell_e^2$ and of the number of steps $s_0$ of the equivalent isotropic random walk. As one can see in Fig.~\ref{sqlealpha},  when $m_0\geq m$, the step length (in units of $\ell^2$) decreases monotonically with increasing $\alpha$. This reflects the fact that collisions become more dispersive with increasing $\alpha$, since deflection angles get larger when the collisions become less dissipative (we already discussed this point in Sec. \ref{sec3}).

We also see in Fig.~\ref{sqlealpha} that $\ell_e^2$ increases with increasing mass ratio $m_0/m$. This is the expected behaviour, since the deflection of intruders due to their collisions with gas particles is less pronounced (and the intruders' motion more persistent) when the mass of the intruder is increased with respect to the mass of the gas particles. Figure \ref{sqlealpha} also highlights the strong increase of $\ell_e^2$ with the mass ratio for very strong inelasticities.

On the other hand, for not too strong inelasticity ($\al\gtrsim 0.8$, say), the number of steps $s_0$ increases very fast with increasing $\al$ (see Fig. \ref{s0alpha}). In contrast, its $\al$-dependence becomes significantly weaker at smaller $\al$-values.

The net effect is that, in the region of small $\al$-values (extreme inelasticity), the (scaled) MSD $R^*= s_0(t^*) (\ell_e^2/\ell^2)$ increases when $\alpha$ decreases; the fastest increase corresponds to the highest mass ratio considered. It is essentially due to the steep behaviour of $\ell_e^2$ in this region. In contrast, for large enough values of $\al$ (nearly elastic spheres), the behaviour of the MSD is dominated by the steep increase of $s_0(t)$, which overcomes the relatively weak decrease in the effective step length as $\alpha$ increases. This competition between the effects of $\alpha$ on both $\ell_e^2$ and $s_0$ and the resulting dependence of the MSD on inelasticity turns out to be quite similar to the one discussed in Sec. \ref{sec3} for the self-diffusive case.

\begin{figure}
\begin{center}
\includegraphics[width=.7\columnwidth]{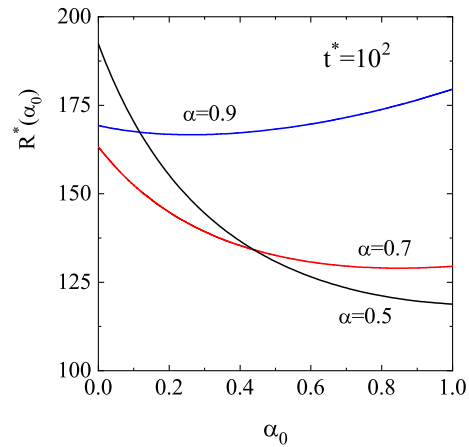}
\end{center}
\caption{Plot of the MSD scaled with the squared mean free path $R^*\equiv \langle |\Delta \mathbf{r}|^2\rangle/\ell^2$ as a function of the coefficient of normal restitution $\al_0$ for $m_0/m=\sigma_0/\sigma=1$, $t^*=10^2$, and three different values of $\al$.
\label{fig5}}
\end{figure}

Having studied the impact of changes in the mass ratio $m_0/m$ on the MSD, we now turn to further assessing the effect of inelastic collisions on this quantity. To study this aspect, we set $m=m_0$, $\sigma=\sigma_0$ \emph{but} $\al \neq \al_0$ (of course, if one takes $\al \neq \al_0$, the self-diffusion results discussed above are recovered). The difference between  $\al $ and $ \al_0$ entails that there is no energy equipartition here either ($T_0/T\neq 1$). This setting is actually relevant for the analysis of inelasticity-driven segregation \cite{SGNT06,SNTG09,BEGS08,BS09}.
Figure \ref{fig5} shows $R^*(\al_0)$-curves for three different values of $\al$ at $t^*=10^5$. A detailed calculation of the scaled MSD $R^*$ as a function of $\alpha_0$ reveals that it decreases monotonically for $\alpha\lesssim 0.6$, and that it increases monotonically for $\alpha\gtrsim 0.94$, whereas in the intermediate range $0.6\lesssim \alpha \lesssim 0.94$,  $R^*(\alpha_0)$ first drops and then raises. We also find that, for a given value of $\al_0$, the scaled MSD decreases with decreasing $\al$ except in a small region with $\al_0$-values around 0.1 or smaller. In this region, the curve for $\al=0.5$ lies above its counterpart for $\al=0.7$.

\vicente{

\section{MSD in driven granular gases}
\label{driven}

In the previous sections, we have analyzed the problem of tracer diffusion in a homogeneously cooling granular gas. However, it goes without saying that the HCS is a rather idealized state, since in real situations one has to externally inject energy into the system to keep it under rapid flow conditions. When the energy added to the granular gas compensates for the energy lost by collisions, a non-equilibrium steady state is attained.

In real experiments, the steady state can be e.g. reached by injecting energy at the system boundaries (which includes shearing or vibrating its walls \cite{YHCMW02,HYCMW04}), but also by bulk driving as in air-fluidized beds \cite{SGS05,AD06}, or by the action of an interstitial fluid \cite{MLNJ01,Yan03,WZXS08}. In practice, these ways of supplying energy involve almost unavoidably the formation of strong spatial gradients in the bulk, thereby driving the system beyond the regime of validity of the Navier-Stokes hydrodynamic description. To circumvent this difficulty, it is quite usual in computer simulations to homogeneously heat the system by the action of an external driving force or \emph{thermostat}. However, thermostats do not play a neutral role in the dynamics; the thermostat indeed modifies the transport coefficients of the granular gas, and it does so in a way which depends on the details of the heating mechanism at hand.

In view of the above, it is of interest to characterize the similarities and differences between the
results derived for the intruder's MSD in the HCS and for their counterparts in the driven granular gas. To this end, we first note that Eq.\ \eqref{12} is still valid for a driven gas. In this case, $s(t)=\nu t$, as $\nu$ no longer depends on time, since the granular temperature $T$ is kept constant by the thermostat. As a result of this, the MSD becomes a linear function of time. In units of the mean free path $\ell$, one has
\beq
\label{dr1}
\frac{\langle |\Delta \mathbf{r} |^2 \rangle}{\ell^2}=d \Bigg[\frac{\Gamma\left(\frac{d}{2}\right)}{\Gamma\left(\frac{d+1}{2}\right)}\Bigg]^2 \widetilde D^* \; t^*,
\eeq
where $t^*=\nu t$, with $\nu$ defined by Eq.\ \eqref{8.1}. The expression of the (dimensionless) diffusion coefficient $\widetilde D^*$ depends on the sort of thermostat used to heat the system. In the framework of our random walk approach, one can easily identify the quantities $s_0(t)$ y $\ell_e^2$ by
setting Eq.\ \eqref{dr1} equal to Eq.\ \eqref{MSDles}. Taking into account that $s_0(t)=\Upsilon t^*$ [with $\Upsilon$ defined by Eq.\ \eqref{Upsilon-def}], one finds
\beq
\label{dr2}
\ell_e^2=d\Bigg[\frac{\Gamma\left(\frac{d}{2}\right)}{\Gamma\left(\frac{d+1}{2}\right)}\Bigg]^2
\left(\frac{\sigma}{\overline{\sigma}}\right)^{d-1}\frac{\chi}{\chi_0}\left(\frac{2\theta}{1+\theta}\right)^{1/2} \widetilde D^* \ell^2.
\eeq
}

\begin{figure}
\begin{center}
\includegraphics[width=.7\columnwidth]{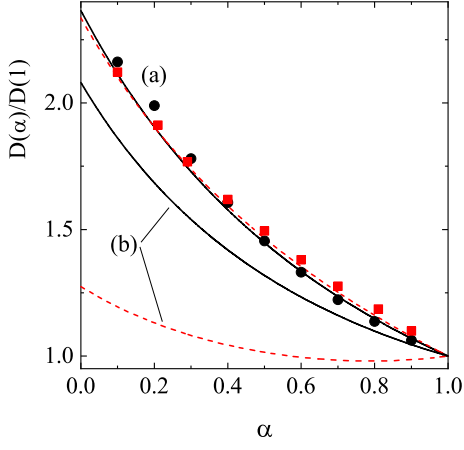}
\end{center}
\caption{Plot of the scaled diffusion coefficient $D(\al)/D(1)$ as a function of the (common) coefficient of normal restitution $\al=\al_0$ for two different systems: $m_0/m=\sigma_0/\sigma=2$ (a) and $m_0/m=0.5$, $\sigma_0/\sigma=1$ (b). In both cases, the solid volume fraction $\phi=0.2$. Here, $D(1)$ refers to the value of the diffusion coefficient for elastic collisions. The solid lines correspond to the results obtained when the granular gas is driven by the stochastic thermostat, while the dashed lines refer to the results obtained in the HCS. The symbols correspond to the DSMC results reported in Refs. \ \cite{GF09} and \cite{GF12} for undriven (squares) and driven (circles) granular gases, respectively.
\label{fig8}}
\end{figure}

\vicente{
Equations \eqref{dr1} and \eqref{dr2} hold for any sort of thermostat.
The thermostat only changes the explicit dependence of the temperature ratio $T_0/T$ and the (dimensionless) diffusion coefficient $\widetilde D^*$ on the system parameters.
To illustrate this dependence, we assume that the granular gas is driven by a Langevin force given by a Gaussian white noise with zero mean and finite variance. The particles are randomly kicked between successive collisions under the action of this stochastic force \cite{WM96,NE98}. In the case of a binary granular mixture (tracer plus gas particles),  the covariance of the stochastic acceleration is chosen to be the same for both species \cite{BT02,G08,G09}.
In the first Sonine approximation, the expression of $\widetilde D^*$ when the system is driven by the stochastic thermostat is \cite{G08,G09}
\beq
\label{dr3}
\widetilde D^*=\frac{1}{\theta \nu_D^*},
\eeq
where $\nu_D^*$ is given by Eq.\ \eqref{17}, and the temperature ratio is obtained from the condition
\beq
\label{dr4}
\frac{m_0}{m}\zeta^*=\frac{T_0}{T}\zeta_0^*,
\eeq
where $\zeta^*$  and $\zeta_0^*$ are given by Eqs.\ \eqref{18} and \eqref{21}. The relation
\ \eqref{dr4} is a cubic equation for $\theta$. As in Sec. \ref{sec3}, the physically meaningful root is
the one which yields the correct value of $\theta=m_0/m$ in the limit of elastic collisions.
}

\begin{figure}
\begin{center}
\includegraphics[width=.7\columnwidth]{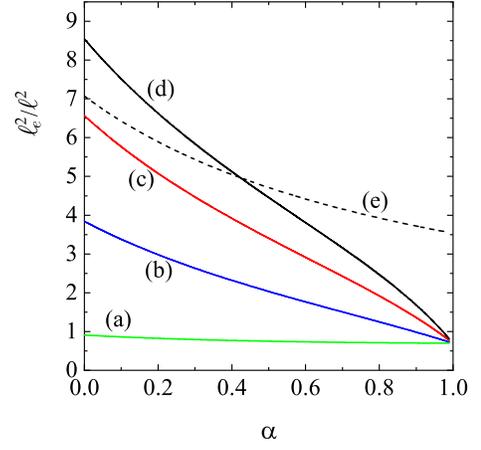}
\end{center}
\caption{Plot of $\ell_e^2/\ell^2$ versus $\al=\al_0$ for a driven granular gas. The solid lines correspond to the cases $m_0/m=0.5$ (a), $m_0/m=5$ (b), $m_0/m=8$ (c), and $m_0/m=10$ (d). In all four cases, the diameter ratio is $\sigma_0/\sigma=2$. The dashed line (e) corresponds to the  self-diffusion case with $m_0/m=1$ and $\sigma_0/\sigma=1$.
\label{fig9}}
\end{figure}

\vicente{Figure \ref{fig8} shows the dependence of the (scaled) diffusion coefficient $D(\al)/D(1)=\widetilde D^*(\al)/\widetilde D^*(1)$ on the (common) coefficient of normal restitution $\al=\al_0$ for $d=3$ and two different mixtures: $m_0/m=\sigma_0/\sigma=2$ (a) and $m_0/m=0.5$ and $\sigma_0/\sigma=1$. Here, $D(1)$ refers to the diffusion coefficient for elastic collisions. In both cases, the solid volume fraction is $\phi\equiv (\pi/6)n \sigma^3=0.2$. For the present case of hard spheres,  $\chi$ is well approximated by the following expression \cite{CS69}:
\beq
\label{dr5}
\chi=\frac{1-\frac{1}{2}\phi}{\left(1-\phi\right)^2}.
\eeq
Similarly, in the case of $\chi_0$ one has \cite{GH72}
\beq
\label{dr6}
\chi_0=\frac{1}{1-\phi}+3 \frac{\sigma_0}{\sigma+\sigma_0}\frac{\phi}{(1-\phi)^2}+2 \left(\frac{\sigma_0}{\sigma+\sigma_0}\right)^2 \frac{\phi^2}{(1-\phi)^3}.
\eeq
It is quite apparent from Fig.\ \ref{fig8} that the influence of the thermostat on $D$ is very weak in the particular case $m_0/m=\sigma_0/\sigma=2$, since the HCS results practically coincide with those for
the thermostatted system. This agreement between both approaches is clearly confirmed by the DSMC results, even for quite extreme values of inelasticity.

However, when $m_0/m=0.5$ and $\sigma_0/\sigma=1$, significant quantitative differences between the theoretical results for the HCS and for the thermostatted system arise. This is the expected result, since in general the impact of the heating mechanism on transport properties is important. In particular, the effect of inelasticity on the intruder's diffusion coefficient is more relevant in the HCS than in the case of a driven gas. We also observe that, in the latter case, the diffusion coefficient exhibits a \emph{monotonic} dependence on $\al$. Thus, according to Eq.\ \eqref{dr1}, the intruder's MSD becomes a monotonic function of the coefficient of restitution. This dependence contrasts
with the non-monotonic behaviour found in the HCS.

The above conclusion (valid for the stochastic thermostat) cannot be generalized to other types of thermostats; in particular, when the granular fluid is driven by a stochastic bath with friction, Sarracino \emph{et al.} \cite{Sarracino10} find a non-monotonic dependence of the diffusion coefficient on $\al$  in the Brownian limit  \cite{KG13}. Consequently, the MSD exhibits a non-monotonic dependence on $\al$, as it is found for the HCS.
Exploring the possible relationship between the HCS [in the stationary representation defined by the scaled time $s(t)$] and the driven case studied by Sarracino \emph{et al.}  requires a more detailed study.

As a complement to Fig.\ \ref{fig8}, in Fig.\ \ref{fig9} we plot the (scaled) square effective length $\ell_e^2/\ell^2$ against the (common) coefficient of restitution $\al=\al_0$ for the case in which the granular gas is driven by the stochastic thermostat. We see that the qualitative dependence of $\ell_e^2/\ell^2$ on $\al$ is quite similar to that found in the freely cooling case (cf. Fig.\ \ref{sqlealpha}). From a quantitative point of view, the influence of inelasticity on the effective length is found to be more pronounced in the undriven case than in the driven one.
}

\section{Conclusions}
\label{Conc}

In summary, starting from the Enskog--Boltzmann kinetic theory, we have derived a general expression for the MSD of an intruder immersed in a granular gas. Although intruder and gas particles are in general assumed to be mechanically different, our approach includes self-diffusion as an interesting special case. The MSD has been found to grow logarithmically in time for any choice of the system parameters; in this sense, the time dependence is very robust and can be ascribed solely to the collisional properties of the gas particles. Homogeneously cooling granular gases were already known to be examples of ultraslow anomalous diffusion; in this context, the self-diffusion case \cite{MJChB14} and the Brownian limit \cite{BRGD99} have been well characterized. However, to the best of our knowledge, no explicit expression valid over the full parameter range has been given so far for the MSD, nor has the behaviour of this quantity as a function of the mass ratio $m_0/m$, the diameter ratio $\sigma_0/\sigma$, and the respective coefficients of normal restitution $\alpha$ and $\alpha_0$ been discussed comprehensively as in the present work.

As we have seen, the prefactor of the logarithmic function exhibits a nonlinear dependence on the characteristic system parameters (masses and diameters, and coefficients of normal restitution). For self-diffusion (namely, when intruder and gas particles are mechanically identical), our results agree qualitatively with results recently obtained by Blumenfeld in the framework of a random walk model \cite{B21}. This agreement is quite remarkable given the minimal amount of assumptions involved in Blumenfeld's model, and may retrospectively be regarded as further evidence supporting the robustness of the logarithmic time growth of the MSD. Nevertheless, the quantitative differences between both results illustrated by Figs.\ \ref{fig1} and \ref{fig2} may become relevant for comparison with both simulation and experiments.

We are not aware of any experimental work where the logarithmic time dependence of the MSD has been confirmed, but a microgravity set-up similar to that of Ref. \cite{YSchSp20} might prove suitable for this purpose. Admittedly, any real experiment will face the inevitable challenge of dealing with the nonequivalence of ensemble- and time-averaged MSD exhibited by the ultraslow diffusion process under study. In principle, this is an important limitation when the number of available trajectories is small, which is the most common situation in experiments. A recent analysis suggests that, for the class of diffusion processes studied here, ergodicity breaking is strongly mitigated at lag times comparable with the trajectory length \cite{BChM15a}. Unfortunately, this is precisely the regime where the shortage of statistics due to the limited number of windows involved in the computation of the time average is most severe.

In this context, we note that explicit use of Haff's law \eqref{2} in Eq. \ \eqref{14}  leads to an exponential relation between the granular temperature of the gas and the intruder's MSD, $T=T(0)\,\exp{(-R^*/C)}$, where $C$ is the prefactor of the logarithm in Eq. \ \eqref{14}. Thus, the intruder's MSD may in principle be used as a ``thermometer'' for the whole granular gas. While this parameter has the advantage of involving the motion of only a single particle, the drawback is the aforementioned typical limitation in the number of trajectories available in experiments. On the other hand, one concludes from the linear theory of error propagation that the error $\delta T$ arising from any uncertainty $\delta R^*$ in the MSD will be strongly dampened by the negative exponential.

 One of the most interesting features borne out by our analysis is the non-monotonic
$\alpha$-dependence of the MSD. This behaviour (not present in Blumenfeld's model) can be understood as a trade-off between two competing effects; on the one hand, the decrease of the collision frequency with decreasing elasticity entails an increase of the MSD, as each collision is a source of antipersistence; on the other hand, decreasing
$\alpha$ results in a smaller root-mean-square velocity, which is detrimental to transport. For $\alpha<\alpha_{min}$, the first effect has been found to prevail over the second one.

The aforementioned non-monotonic $\alpha$-dependence of the MSD at sufficiently long times can also be rationalized by considering a random walk with isotropic and uncorrelated steps that is equivalent to the actual one in the sense that it yields the same MSD. This MSD can be written as a product of the square effective step length $\ell_e^2$ and the step number $s_0$. The first quantity decreases monotonically with increasing $\alpha$, but the second one increases strongly with $\alpha$ for large enough values of this parameter.

Finally, for mechanically different intruders with $\sigma_0/\sigma=m_0/m=1$, we see that the MSD behaves non-monotonically both as a function of $\alpha_0$ for a fixed
$\alpha$ and as a function of $\alpha$ for a given $\alpha_0$.
 In contrast, the $m_0/m$-dependence of the MSD has been found to be monotonic in the explored parameter range ($m_0>m$). The fact that one can minimize the mobility of an intruder after a given time by a convenient choice of its coefficient of restitution is not particularly intuitive and awaits experimental confirmation. \vicente{ As we have seen, this finding does not carry over to all driven systems, since
 the dependence of the MSD on $\alpha$ can be monotonic or non-monotonic depending on the kind of
 thermostat chosen. In this context, studying the behaviour of the velocity correlations
 for both driven \cite{FAZ09} and freely cooling systems \cite{DBL02,AP07,HO07} might shed further light on some salient features of the observed phenomenology, such as backscattering and aging effects.}

\textbf{Acknowledgments}

We thank R. Blumenfeld for  stimulating discussions. We acknowledge financial support from Grant\\
PID2020-112936GB-I00 funded by MCIN/AEI/\\
10.13039/501100011033, and from Grants IB20079, \\ GR18079 and GR21014 funded by Junta de Extremadura (Spain) and by ERDF A way of making Europe.

\vspace{0.5cm}

\textbf{Declarations}\\

\textbf{Conflict of interest.} The authors declare that they have no conflict of interest.

\appendix
\section{Collision frequency of the intruder}
\label{appendixA}

For hard spheres, the (average) collision frequency $\nu_0$ associated with the intruder is defined as
\beqa
\label{a1}
\nu_0(t)&=&n_0^{-1}\overline{\sigma}^{d-1}\chi_0\int d\mathbf {v}_1 \int d\mathbf{v}_{2}\int d\widehat{\boldsymbol{\sigma}}\,\Theta (\widehat{{\boldsymbol {\sigma}}}\cdot {\bf g}_{12})(\widehat{\boldsymbol {\sigma }}\cdot {\bf g}_{12})\nonumber\\
& & \times f(\mathbf{v}_1;t)f_0(\mathbf{v}_2;t),
\eeqa
where $f(\mathbf{v}_1;t)$ and $f_0(\mathbf{v}_2;t)$ are the velocity distribution functions of the granular gas and the intruders, respectively. In addition, $\widehat{\boldsymbol {\sigma}}$ is a unit vector along the line of the centers of the two spheres at contact, $\Theta $ is the Heaviside step function, and ${\bf g}_{12}={\bf v}_{1}-{\bf v}_{2}$ is the relative velocity. The integrals appearing in Eq.\ \eqref{a1} are evaluated here by considering the Maxwellian approximations to $f$ and $f_0$, namely,
\beq
\label{a2}
f(\mathbf{v}_1)\to n \pi^{-d/2} v_\text{th}^{-d} e^{-c_1^2}, \quad f_0(\mathbf{v}_2)\to n_0 \pi^{-d/2} \theta^{d/2} v_\text{th}^{-d} e^{-\theta c_2^2},
\eeq
where $v_\text{th}=\sqrt{2T/m}$, $\mathbf{c}_i=\mathbf{v}_i/v_\text{th}$, and $\theta=m_0T/(mT_0)$. Thus, $\nu_0$ can be rewritten as
\beq
\label{a3}
\nu_0(t)=n\overline{\sigma}^{d-1}\chi_0  \theta^{d/2} v_\text{th}(t) I_\nu,
\eeq
where we have introduced the dimensionless integral
\beq
\label{a4}
I_\nu=\pi^{-d}\int d\mathbf {c}_1 \int d\mathbf{c}_{2}\int d\widehat{\boldsymbol{\sigma}}\,\Theta (\widehat{{\boldsymbol {\sigma}}}\cdot {\bf g}_{12}^*)(\widehat{\boldsymbol {\sigma }}\cdot {\bf g}_{12}^*)e^{-c_1^2-\theta c_2^2}.
\eeq
Here, $\mathbf{g}_{12}^*=\mathbf{g}_{12}/v_\text{th}$. The integral $I_\nu$ can be performed by the change of variables $\mathbf{x}=\mathbf{c}_1-\mathbf{c}_2$, $\mathbf{y}=\mathbf{c}_1+\theta \mathbf{c}_2$, with the Jacobian $(1+\theta)^{-d}$. The integral $I_\nu$ gives
\beqa
\label{a5}
I_\nu&=& \frac{S_d^2}{\pi^{(d+1)/2}\,\Gamma\left(\frac{d+1}{2}\right)}\,(1+\theta)^{-d}\int_0^\infty dx x^d e^{-ax^2}\nonumber\\
& & \times \int_0^\infty dy y^{d-1} e^{-b y^2},
\eeqa
where $S_d=2\pi^{d/2}/\Gamma(d/2)$ is the total solid angle in $d$ dimensions, $a\equiv \theta(1+\theta)^{-1}$,  $b\equiv (1+\theta)^{-1}$ and use has been made of the result \cite{NE98}
\beq
\label{a6}
 \int d\widehat{\boldsymbol{\sigma}}\,\Theta (\widehat{{\boldsymbol {\sigma}}}\cdot {\bf g}_{12}^*)(\widehat{\boldsymbol {\sigma }}\cdot {\bf g}_{12}^*)=\frac{\pi^{(d-1)/2}}{\Gamma\left(\frac{d+1}{2}\right)} \, \mbox{g}_{12}^*.
\eeq
The integral $I_\nu$ gives
\beq
\label{a7}
I_\nu=\frac{\pi^{(d-1)/2}}{\Gamma\left(\frac{d}{2}\right)}\theta^{-d/2}\left(\frac{1+\theta}{\theta}\right)^{1/2},
\eeq
and hence, $\nu_0$ can be finally written as
\beq
\label{a8}
\nu_0(t)=\frac{\pi^{(d-1)/2}}{\Gamma\left(\frac{d}{2}\right)} n\overline{\sigma}^{d-1}\chi_0 \left(\frac{1+\theta}{\theta}\right)^{1/2} v_\text{th}(t).
\eeq

When the intruder and gas particles are mechanically equivalent, $\overline{\sigma}=\sigma$, $\chi_0=\chi$, $\theta=1$, and Eq.\ \eqref{a8} yields
\beq
\label{a9}
\nu_0(t)=\frac{\sqrt{2}\pi^{(d-1)/2}}{\Gamma\left(\frac{d}{2}\right)} n \sigma^{d-1}\chi v_\text{th}(t).
\eeq
The expression \eqref{a9} agrees with Eq.\ \eqref{8.1}. When the intruder and gas particles are different, Eq.\ \eqref{a8} leads to Eq.\ \eqref{16.0}.

\section{Average of $\ln\left(1+\lambda v_0\right)$ over the initial velocity distribution}
\label{appendixB}

In what follows, we compute explicitly the Boltzmann average of $\ln\left(1+\lambda v_0\right)$, i.e., the integral
\beq
\mathcal{I}\equiv \frac{1}{n_0}\int d\mathbf{v}_0\; \ln(1+\lambda v_0)\; f_\text{M}(\mathbf{v}_0),
\eeq
where $f_\text{M}(\mathbf{v}_0)$ is given by Eq.\ \eqref{M.2}. Introducing now the dimensionless quantities $\hat{\mathbf{v}}_0=\mathbf{v}_0/v_{th}(0)$ and $\hat{\lambda}=v_{th}(0) \lambda$, we get
\beq
\label{avgint}
\mathcal{I}=\int_0^\infty d\hat{v}_0\; \ln\left(1+\hat{\lambda} \hat{v}_0\right) \hat{f}_\text{M}(\hat{v}_0),
\eeq
with
\beq
\hat{f}_\text{M}(\hat{v}_0)=\frac{4}{\sqrt{\pi}}\, \hat{v}_0^2 \,e^{-\hat{v}_0^2}.
\eeq
The integral $\mathcal{I}$ has been computed by using a computer package of symbolic calculation. Its expression is given in terms of the hypergeometric functions. However, since the above expression of $\mathcal{I}$ is quite cumbersome, its explicit form is omitted here.  For large $\hat{\lambda}$ (or, equivalently, large $t$), the corresponding expansion yields a much simpler asymptotic form:
\beq
\mathcal{I}=\left(\frac{\ln(\hat{\lambda})+C_1}{\ln(\hat{\lambda})+C_2}+
{\cal O}(\hat{\lambda}^{-1}\ln^{-1}\hat{\lambda})\right) \ln\left(1+\lambda \bar{v}_0\right).
\eeq
where $C_1=1-\ln(2)-(\gamma_E/2)$, $C_2=\ln(2)-(1/2)\ln(\pi)$, and $\gamma_E\simeq 0.5772$ denotes the Euler-Mascheroni constant. Thus, at long times,
\beq
\mathcal{I} \approx [1-(C_2-C_1)\ln^{-1}(v_{th}(0)\lambda)] \ln\left(1+\lambda \bar{v}_0\right).
\eeq
In other words, the convergence of $\mathcal{I}$ towards  $\ln\left(1+\lambda \bar{v}_0\right)$ in the long-time limit is dictated by an inverse logarithmic law.

\end{document}